\documentclass[doublecol]{epl2} 
\usepackage{amssymb,amsfonts,latexsym,graphicx}
\usepackage{amsmath,bm}
\usepackage[dvipsnames,usenames]{xcolor}
\usepackage{hyperref}
\usepackage{epstopdf}
\renewcommand{\vec}[1]{\boldsymbol{#1}}
\newcommand{\h}{\mathcal{H}} 
\newcommand{\M}{\mathcal{M}} 
\newcommand{\Sx}{ \partial_x S} 
\newcommand{\Sy}{ \partial_y S} 
\newcommand{\St}{ \partial_t S} 
\newcommand{\T}{\hat{T}}

\newcommand{\atil}{\tilde{\alpha}} 
\newcommand{\tc}{\langle t_c \rangle} 

\usepackage{mathbbol}

\title{Emergence and dynamical properties of stochastic branching in the electronic flows of disordered Dirac solids}
\shorttitle{Branching in  disordered Dirac solids} 

\author{Marios Mattheakis\inst{1} \and G. P. Tsironis\inst{1,2} \and Efthimios  Kaxiras\inst{1,3}}
\shortauthor{M. Mattheakis \etal}

\institute{                    
  \inst{1} School of Engineering and Applied Sciences, Harvard %
  University, Cambridge, Massachusetts 02138, USA\\
  \inst{2} Department of Physics, University of Crete, 
  Heraklion 71003, Greece\\
  \inst{3} Department of Physics, Harvard University, Cambridge, %
  Massachusetts 02138, USA
}

\abstract{
Graphene as well as more generally Dirac solids constitute  two dimensional materials where the electronic flow is ultra relativistic.  When a Dirac solid is deposited on a different substrate 
surface with roughness, a local random potential develops through an inhomogeneous charge impurity 
distribution.  This external potential affects profoundly the charge flow and induces a chaotic pattern of current branches that develops through focusing and defocusing effects produced by the randomness of the surface.  An additional bias voltage may be used to tune the branching pattern of the charge carrier currents.  We employ analytical and numerical techniques in order to investigate the onset and the statistical properties of carrier branches in Dirac solids.  We find a specific scaling-type relationship that connects the physical scale for the occurrence of branches with the characteristic medium properties, such as disorder and bias field.  We use numerics to test and verify the theoretical prediction as well as a perturbative approach that gives a clear indication of the regime of validity of the approach.  This work is relevant to device applications and may be tested experimentally.}

\begin{document}

\maketitle

\section{Introduction}  Wave focusing due to refractive index variation is a common occurrence in many physical systems. In sea waves, the effective refractive index variation arises from fluctuating depth
\cite{natPhys12_2016, natPhys12_2016_heller, jfluid790_2016, prl118_2017}; in optical \cite{wavesRandom3_1993, chaos22_2012, prl111_2013, myBranchingChapter, myRWs_2016, prl119_2017} or other media \cite{acoustic107_2000, acoustic109_2001} the index of refraction changes in a statistical way due to small imperfections or distributions of defects. Random  spatial variability  of the index leads to local focusing and defocusing of the waves and the formation of {caustics} or wave {branches} with substantially increased  local wave intensity.  Under general circumstances the branching flow develops a stochastic web with   statistical patterns of persisting enhanced intensity wave motion.  Since electrons have also wave properties due to their  quantum nature, similar phenomena  appear in the quantum realm. Injected electrons in disordered two dimensional (2D) electron gas form coalescing trajectories and manifest phenomena similar to wave motion in random media \cite{nature410_2001, prl89_2002, natPhys3_2007, prl105_2010}.  One aspect of the electronic motion that has not yet been explored is the relativistic and in particular the {ultra-}relativistic one; the latter  occurs in materials referred to as {Dirac solids} (DS), such as graphene \cite{nature438_2005, advancePhys63_2014, prb93_2016,nanoLet16_2016,science351_2016}. {In the ultra-relativistic limit, the magnitude of  the Dirac fermion velocity cannot be
 affected by external fields as it is already at its maximum value, leading to significant differences from the conventional  non-relativistic flow. This   is directly reflected  in the electronic branching properties and gives rise to discernible differences, as we show here. }

Pristine graphene is the prototypical 2D-DS, characterized by linear dispersion in the electronic band structure   near the Fermi level \cite{nature438_2005, advancePhys63_2014},
\begin{equation}
\label{eq:ek}
\epsilon_{\bf k} = v_F \hbar |{\bf k}|,
\end{equation}
where $\epsilon_{\bf k}$ is the single particle energy, $v_F$ is the Fermi velocity and ${\bf k}$ is the wave-vector.  Electron flow in these band segments is ultra-relativistic with maximal propagation velocity $v_F$  \cite{nature438_2005,  science351_2016, epjb85_2012,pre87_2013, natPhys4_2008}. This electron flow is  modified by the presence of a bias potential  applied  along a specific direction.  The relativistic electronic dispersion couples the motion of electrons along  the direction of bias and the one perpendicular to it \cite{epjb85_2012, pre87_2013}.  Electron  dynamics is also subject to the presence of substitutional or other type of weak disorder; the effects of such disorder  are  observed in graphene in the form of electronic puddles \cite{prl99_2007, natPhys4_2008,  natPhys5_2009, prb91_2015, prl116_2016}. The combined presence of disorder and bias  alters the electronic flow and thus the ultra-relativistic  trajectories coalesce into branches of substantial local density.  In this letter we show that the weak surface disorder  produces a lensing mechanism for the electronic waves that is  clearly manifested in the form of caustics. 
Dirac fermions cannot be accelerated by a bias potential, in contrast to the corresponding behavior of non-relativistic electrons in 2D metals with parabolic bands.  As a result, a simple relationship derived here between the caustic location and statistical properties of the intrinsic  potential  remains valid  even in the presence of  external fields; we derive this scaling-type formula  analytically and  verify it by numerical simulations.
Scanning probe microscope (STM) techniques have been  used extensively to measure the electron density \cite{nanoLet16_2016} and in turn, the branching flow in 2D electron gases \cite{nature410_2001,natPhys3_2007}.

\section{Theoretical Analysis} The basic feature of a 2D-DS is the linear dispersion relation of the energy with wave-vector, Eq. (\ref{eq:ek}). When the electronic density is low we may use the independent electron model to describe the quasi-classical dynamics  of charge carriers {\cite{epjb85_2012, pre87_2013} through the quasi-classical ultra-relativistic Hamiltonian 
\begin{equation}
\label{eq:relativisticHamiltonian}
\h = \pm v_F\sqrt{p_x^2 + p_y^2} +V(x,y),
\end{equation}
where   $\vec{p} = (p_x, p_y)$ is the 2D momentum of the  charge carriers. 
The  Hamiltonian  (\ref{eq:relativisticHamiltonian}) is the classical limit  of Dirac equation \cite{epjb85_2012} that describes dynamics of massless electron/hole quasi-particles in graphene and other DS \cite{advancePhys63_2014}.
Branched flow is an effect of  ray fields associated with waves \cite{natPhys12_2016, acoustic107_2000,  prl118_2017} and thus, we may use the  Hamiltonian (\ref{eq:relativisticHamiltonian}) to give a ray description of the quantum flow of Dirac electrons. 
{We note that  STM measurements  have shown that classical ray-tracing simulations describe accurately the electron flow in graphene \cite{nanoLet16_2016}.}

We focus on 2D ultra-relativistic dynamics of particles in a medium with   potential  $V(x,y) = V_r (x,y) + V_d (x)$,  where  $V_r (x,y)$  is a  random $\delta$-correlated potential with  energy scale much smaller than that of the  electronic flow; this term comes  from random charged impurities in the substrate or other sources of disorder in the graphene sheet \cite{natPhys4_2008, natPhys5_2009, prl99_2007, prb91_2015, prl116_2016}. The  deterministic ``control''  potential $V_d (x)=-\alpha x$  is due to an externally applied voltage in the $x$-direction with $\alpha$  fixed by experimental conditions.
Electrons are injected in the Dirac  sheet with  initial momentum $p_0$ along the $x$-direction; due to the ballistic electronic motion along the $x$-axis we may ignore the effects of the random potential in this direction  \cite{prl89_2002, prl105_2010}. We use plane wave initial conditions for the electrons,    $p_x(0)=p_0$  and $p_y(0)=0$,  and write the corresponding solution of  Hamilton's equations   as $p_x(t)=p_0+\alpha t$ and $p_y(t)=-\int{\partial_y V_r(x,y) dt}$, with  $p_x\gg p_y$.     We expand  the Hamiltonian (\ref{eq:relativisticHamiltonian}) up to second order in $p_y/p_x$ to obtain 
 
\begin{equation}
\label{eq:H}
\h = p_x + \frac{p_y^2 }{2 p_x} +V(x,y),
\end{equation}
and set $v_F=p_0=1$ for simplicity.
In order to study wave-like electronic flow through the Hamilton-Jacobi equation (HJE) we  introduce the classical action $S$  with  $p_x =\Sx $ and $p_y=\Sy$, and express the HJE as  
\begin{equation}
\label{eq:HJE}
\St + \Sx +\frac{ (\Sy)^2 }{2 \Sx} +V(x,y) =0.
\end{equation}
Employing previous non-relativistic approaches  to the branching problem we derive an  equation for the local curvature of the electronic flow determined through $u=\partial_{yy}S$ \cite{wavesRandom3_1993, myBranchingChapter}.  To this effect we  apply the operator $\T\equiv (\partial_{xx} +\partial_{yy} +2\partial_{xy} )$ on Eq.
 (\ref{eq:HJE})  and obtain
\begin{equation}
\label{eq:curvature1}
\partial_t u + \frac{u^2}{\Sx} + \frac{\Sy}{\Sx}\partial_y u + \T V(x,y) =0.
\end{equation}
Using the effective Hamiltonian (\ref{eq:H}) we calculate the  equations of motion for the $x$, $p_x$ conjugate variables keeping   the lowest order terms in $p_y/p_x$, which leads to the simple relation  $ x(t) = t$. Thus, under the approximation of the dominance of the momentum in the forward $x$-direction, we find that space and time variables are identical.  We use  this fact to simplify the HJE ray dynamics and turn  Eq. (\ref{eq:curvature1}) into a quasi-2D version \cite{natPhys12_2016, wavesRandom3_1993}, that is, we replace the space coordinate $x$ with  $t$ and use  $\T V = \partial_{yy}V_r(t,y)$; this simplification is valid only in the ultra-relativistic limit since non-relativistic electrons accelerate in the presence of non-zero values for the parameter $\alpha$. 

We need to calculate the convectional derivative for $u$; for an arbitrary 
 function $f(\h)$, where $\h$ is the Hamiltonian (\ref{eq:H}) we have 
 \begin{equation}
\label{dfdt}
\frac{d f}{dt} = \left[ {\partial_t} + \frac{\partial x}{\partial t} {\partial_x} + \frac{\partial y}{\partial t} {\partial_y}  \right]f.
\end{equation}
In quasi-2D approximation the potential depends only on time and transverse coordinate $y$, i.e.  $V \equiv V(t,y)$; subsequently, the term that includes the operator ${\partial_x}$ is zero.  On the other hand, the term  ${\partial y}/{\partial t}$ can be determined  by  Hamilton's equations of (\ref{eq:H}) as
\begin{equation}
\label{eq:dydt}
\frac{\partial y}{\partial t} = \frac{d y}{dt} = \frac{\partial \h}{\partial p_y} = \frac{p_y}{p_x}
\end{equation}
Using the definition of the classical action $S$, namely $p_x = \partial_x S$ and $p_y = \partial_y S$, and  the Eqs. (\ref{dfdt}) and  (\ref{eq:dydt}) we obtain the convectional derivative formula:
\begin{equation}
\label{eq:dt}
\frac{d f}{dt} = \left[ {\partial_t} + \frac{\partial_y S }{\partial_x S} {\partial_y}  \right]f.
\end{equation}
Using the expression of Eq. (\ref{eq:dt})  for the convectional derivative in conjunction with the approximate quasi-one dimensional Hamiltonian of Eq. (\ref{eq:H}) we obtain an
ordinary nonlinear differential equation for the local wave curvature:
\begin{equation}
\label{eq:ODEu}
\frac{d u}{dt} + \frac{u^2}{1+\alpha t} +\partial_{yy}V_r(t,y) = 0.
\end{equation}
The dynamics of Eq. (\ref{eq:ODEu}) determines the onset of the regime for caustics; this occurs at times, or equivalently   locations along the $x$-axis, where  the curvature $u$ becomes singular \cite{wavesRandom3_1993, prl89_2002, prl105_2010}.  The first time when a singularity in $u$ occurs  determines the precise point for the onset of  ray coalescence.   Given that the term $ \partial_{yy}V_r(t,y)$ is fluctuating, we  solve  the  first passage time problem for the curvature to reach  $|u(t_c)|\rightarrow  \infty$,  where $t_c$ is the  time for the occurrence of the first caustic.

\section{Deterministic caustic dynamics} We may obtain a useful and intuitive expression for the onset of branches if we ignore at first the  stochastic potential term of Eq. (\ref{eq:ODEu}); straightforward solution of the resulting simple nonlinear equation leads to the solution 
\begin{equation}
\label{eq:tc0}
 t_c = \frac{e^{\alpha/|u_0 |}-1}{\alpha},
\end{equation}
where we set $u_0 =-|u_0 |$ since negative initial curvature leads to positive $t_c$. In the  strong external potential limit $(\alpha\rightarrow \infty)$ the caustic needs infinite time to develop    ($t_c \rightarrow \infty$), while in the weak  limit of    $(\alpha \rightarrow 0)$  $t_c$ is finite and increases  linearly with $\alpha$,
\begin{equation}
\label{eq:tc_a}
t_c = \frac{1}{|u_0|}\left( 1 + \frac{\alpha}{2 |u_0|}\right).
\end{equation} 
This behavior follows from the effective elimination of the nonlinear term in $u$  of Eq. (\ref{eq:ODEu}),  in the large $\alpha$ limit, which is  responsible for the  creation of caustic events.

Along the  propagation axis $t$, the fluctuating term in Eq. (\ref{eq:ODEu})  acts as a $\delta$-correlated noise $\xi(t)$ with zero mean  and  standard deviation $\sigma$,    $\partial_{yy} V_r(t,y)=\sigma^2 \xi(t)$, and  Eq. (\ref{eq:ODEu}) becomes a stochastic Langevin equation \cite{prl118_2017, prl89_2002, prl105_2010}. For $\alpha=0$ the curvature Eq. (\ref{eq:ODEu}) reduces to the non-relativistic case with the average first caustic time $\langle t_c \rangle$ obeying the scaling relationship $\langle t_c \rangle \sim \sigma^{-2/3}$  \cite{natPhys12_2016, wavesRandom3_1993, prl105_2010}.

\section{Self-consistent equation for the onset of branches} To quantify the location of the occurrence of  the  first relativistic caustic event  including $\alpha$, we use a self-consistent approach.  Specifically, we start from the more general equation
\begin{equation}
\label{eq:odeu}
\frac{d u}{dt} = - \frac{u^2}{p_0+\alpha (t-t_0)} -\sigma^2 \xi(t),
\end{equation}
where   $\alpha$ is the deterministic control parameter, $\sigma$ is the standard deviation of random potential, and $\xi$ a white noise with unit variance. To find self-consistently a new scaling-type relationship we assume $t=\tc$ turning   Eq. (\ref{eq:odeu})  to  
\begin{equation}
\label{eq:ODEuGamma}
\frac{d u}{dt} = -\gamma u^2 -\sigma^2 \xi(t),
\end{equation}
where $\gamma$ is a constant defined as
\begin{equation}
\label{eq:gamma}
\gamma = \frac{1}{ p_0-\alpha t_0  +  \alpha  \langle t_c \rangle}.
\end{equation} 
The Eq. (\ref{eq:ODEuGamma}) has been treated in refs. \cite{wavesRandom3_1993,myBranchingChapter}   for the simpler case of
$\gamma=1$ yielding $\tc =3.32 \sigma^{-2/3}$. 
Following the same stochastic approach for $\gamma \neq 1$, we obtain the more general formula
\begin{equation}
\label{eq:tc1}
\langle t_c \rangle = 3.32 (\sigma \gamma)^{-2/3}.
\end{equation}

The new expression is derived after the solution  of the self-consistent set of Eqs. (\ref{eq:gamma}) and (\ref{eq:tc1}); the combined equation is given by
\begin{equation}
\label{eq:3dpolynomial}
A x^3 + B x^2 + 1 =0
\end{equation}
where 
\begin{equation}
\label{eq:ABx}
A= -\frac{3.32^{-3/2}}{ p_0-\alpha t_0 }\sigma, \quad B=\frac{\alpha}{p_0-\alpha t_0}, \quad x=\tc^{1/2}.
\end{equation}
The only real solution of the Eq. (\ref{eq:3dpolynomial}) is 
\begin{equation}
\label{eq:x}
x = \frac{B}{3A}\left(1+ \frac{  B}{ c} +\frac{ c}{  B}  \right),
\end{equation}
where
\begin{equation}
\label{eq:c}
c  = \left[ \frac{27 A^{2}}{2} \left(1+ \frac{2 B^3}{27 A^2} +\sqrt{1+\frac{4 B^3}{27 A^2}} \right)\right]^{1/3}.
\end{equation}
Since $\alpha$ and $\sigma$ are assumed to have small values,  we define the small coefficient
\begin{equation}
\label{eq:delta}
\atil = \frac{ B A^{-2/3}}{ 3 } \ll 1.
\end{equation}
Substitution of Eq. (\ref{eq:delta}) in  (\ref{eq:c}) and expanding  for small $\atil$  yields approximately to
\begin{equation}
\label{eq:beta}
c = 3 A^{2/3} \left(1 + \frac{2}{3}\atil^3 \right).
\end{equation}
We use  Eq. (\ref{eq:beta}) in  (\ref{eq:x}) and expand up to $\atil^3$  to obtain
\begin{equation}
\label{eq:x2}
x = -A^{-1/3} \left(1 + \atil +\atil^2 +  \frac{2}{3}\atil^3 \right).
\end{equation}
Using Eqs. (\ref{eq:ABx})  and  (\ref{eq:x2}), we find the ultra-relativistic formula for the first caustic time
\begin{equation}
\label{eq:tcRelativistic}
\tc = 3.32 \left(\frac{\sigma}{p_0-\alpha t_0}\right)^{-2/3} \left(1 +2 \tilde{\alpha} +3 \tilde{\alpha}^2+\frac{10}{3}\tilde{\alpha}^3\right)
\end{equation}
with  
\begin{equation}
\tilde{\alpha}=1.11 \frac{\alpha \sigma^{-2/3}}{\left(p_0-\alpha t_0\right) ^{1/3}}  .
\end{equation}

Choosing a more convenient set of initial values, i.e. $t_0=0$ and $p_0=1$, yields
to
\begin{equation}
\label{eq:tc}
\langle t_c \rangle \sim \sigma^{-2/3}\left(1+ 2\tilde{\alpha} + 3\tilde{\alpha}^2 +\frac{10}{3}
\tilde{a}^3 \right),
\end{equation}
with $\tilde{\alpha}= 1.11 \alpha \sigma^{-2/3}$  the  relativistic correction term in the presence of a deterministic potential.
We note that for $\tilde{\alpha}=0$ we obtain the previously obtained expression for not relativistic branches \cite{prl105_2010}.

\section{Numerical solution of the Hamilton-Jacobi Equation}
We now depart from the   quasi-2D approximation and   solve numerically the characteristic equations for the full Hamiltonian (\ref{eq:relativisticHamiltonian}) while constructing a  random potential based on experimental observations. In particular,  impurities in the substrate of  graphene create a smooth landscape of charged puddles of radius $R\approx 4$ nm  \cite{prb91_2015, prl116_2016}.  In our model, each puddle size is drawn from a  two-dimensional Gaussian distribution with   standard deviation $R$.  The location of each puddle is randomly chosen through a uniform distribution. 
We perform simulations for quasi-classical electron dynamics in a graphene sheet of size $400\times400$ nm, where several caustics are observed;  periodic boundary conditions  are used to ensure that all the rays  reach a caustic.  The random potential consists of 2000 randomly distributed Gaussian defects with $R=4$ nm.
A collection of 1000 ultra-relativistic rays,  initially distributed uniformly along $y$ axis, are injected into the graphene sheet from the $x=0$, with plane wave initial conditions,  $p_x(0)=p_0=1$ and $p_y(0)=0$. 
We select a single caustic event out of many  to show   how the  deterministic part of the potential affects the onset of this event, see  Fig. \ref{fig:manycaustics}. 
The rays propagate in the disordered potential  with $\sigma=0.1$ and after time $t_c$ a  caustic {event} occurs. 
The  ray-tracing simulations are performed  for three different values of  $\alpha=[0,~0.05,~0.1]$, to show that $t_c$ increases linearly  with $\alpha$, a behavior  expected from  Eq. (\ref{eq:tc_a}).  We thus confirm numerically the quasi-2D analytical  prediction that the presence of a small voltage in graphene shifts the location of the first caustic, a fact that can be tested experimentally. The location and the shape of caustics are modified by the external potential $V_d$; in particular, as $\alpha$ increases the passage to branched flow is delayed and the caustics disperse slower, see Fig. \ref{fig:manycaustics}.

\begin{figure}[ht]
\includegraphics[scale=.27]{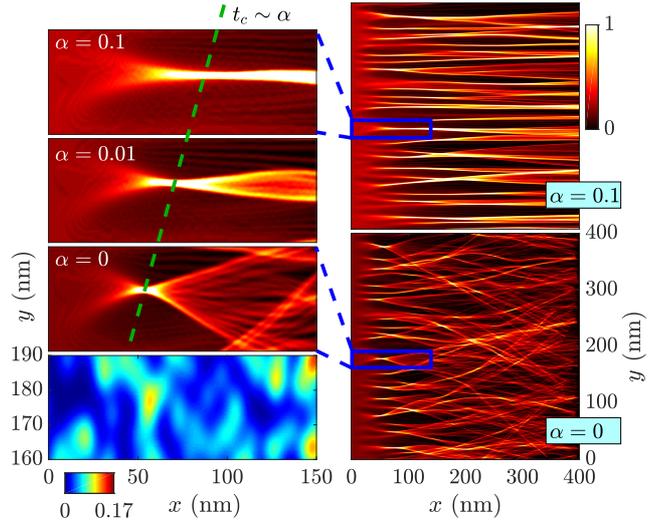}
\caption{Two-dimensional numerical ray-simulations determine the onset of a caustic event in a disordered potential  with $\sigma=0.1$ and for a deterministic potential with $\alpha= [0,~0.05,~0.1]$. (Left) The lower panel  shows the random potential. The remaining images represent the density of rays $I$. The green dashed line shows that the  first caustic time $t_c$ increases linearly with $\alpha$. (Right) The ray density of branched flow in a graphene sheet for $\alpha=0$  and $\alpha=0.1$.} \label{fig:manycaustics}
\end{figure}

\section{Phase space approach} The  classical  electron trajectories can be used to determine the onset of a caustic   in the context of the stability matrix $\M$. The latter  describes the evolution in time of an infinitesimal volume of phase space, $\delta x_i(t)=\M\delta x_i(0)$, where $x_i=(x,y,p_x,p_y)$ is  the four dimensional phase space vector, and the elements of $\M$ read  $m_{ij}(t)=\partial x_i(t)/\partial x_j(0)$ \cite{acoustic109_2001, prl105_2010}. The evolution of $\M$ is given by $\dot \M(t) = \mathbf{K} \M(t)$ with initial condition $m_{ij}(0)=\delta_{ij}$, i.e.
\begin{equation}
\label{eq:dotM}
\dot \M(t) = \left( \begin{array}{cc} 0 &   \mathbb{1}  \\ - \mathbb{1} & 0 \end{array} \right) \frac{\partial^2 \h}{\partial x_i \partial x_j} \M(t)=\mathbf{K} \M(t),
\end{equation}
where $\mathbb{1}$ is the identity matrix. For Dirac fermions the symplectic matrix $\mathbf{K}$ is obtained from the  ultra-relativistic Hamiltonian: 
\begin{equation}
\label{eq:K}
\mathbf{K} =  
\left( \begin{array}{cccc} 0 & 0 & {  p_y^2}/{p^3}  & -{p_x p_y}/{p^3} \\ 
 0 & 0 & -{p_x p_y}/{p^3}  & { p_x^2}/{p^3} \\  -V_{xx} & -V_{xy} & 0  & 0 \\ -V_{xy} & -V_{yy} & 0 & 0 
\end{array} \right),
\end{equation} 
where $p=|{\bf p}|=\sqrt{p_x^2+p_y^2}$.
A caustic occurs when the classical density of rays diverges giving rise to the  condition  for caustic emergence $\left( -p_y,p_x,0,0 \right)^T \M  \delta x_i(0) = 0$ \cite{acoustic109_2001}.  We use this  condition to determine numerically the time $t_c$ where the first caustic occurs by  calculating the average time needed for the first caustic event. To this end  we study  a wide range of strength for the Gaussian defects in order to obtain many $V_r$  with different variances $\sigma^2$. In addition, we examine several bias  potentials with $\alpha$ ranging  between $0$ and $0.1$ while realizing 30 disorder potentials for each pair of $\sigma$ and $\alpha$ values. Furthermore, in these simulations we consider  $10^4$ ultra-relativistic electrons  distributed uniformly along $y$ and ejected at $x=0$.}
In Fig. \ref{fig:scaling} we present the relation between $\langle t_c \rangle$ and the potential parameters $(\sigma, \alpha)$.  For $\alpha=0$ we obtain  $\langle t_c\rangle \sim \sigma^{-2/3}$, the scaling of conventional 2D metals in the absence of bias potential \cite{prl89_2002, prl105_2010}. 
 When $\alpha \neq 0$,   $\ln(\langle t_c \rangle)$ decays practically linearly with $\ln(\sigma)$  revealing that the ultra-relativistic nature of Dirac fermions retains the scaling also in the presence of  a bias potential, as predicted theoretically by Eq. (\ref{eq:tc}).  
{The color solid lines show the theoretical prediction of $\langle t_c \rangle$ through the use of  Eq.  (\ref{eq:tc}) in the range that  is valid,  $\alpha \sigma^{-2/3} \ll 1$.}

\begin{figure}
\centering
\includegraphics[scale=0.31]{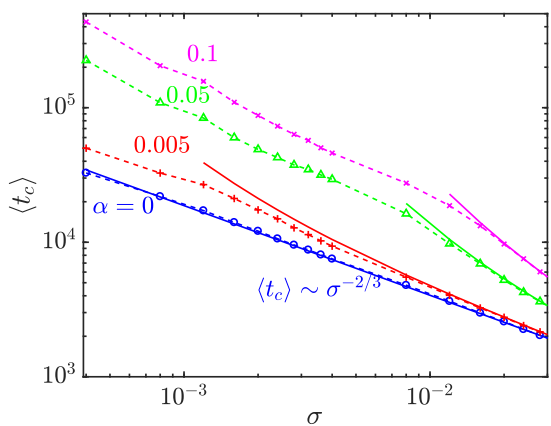} 
\caption{Simulation results for the mean first  caustic time $\langle t_c \rangle$ for the occurence of a caustic event as a function of the random potential standard deviation $\sigma$ and for bias  determined by  $\alpha$ in the range $[0,~0.1]$ (dashed lines conect data points). The color solid  lines indicate the theoretical predicted relationship  between $\langle t_c \rangle$ and $\sigma$ in log space, in the range where Eq. (\ref{eq:tc}) is valid.} \label{fig:scaling}
\end{figure}

\section{Dirac solids with a gap} The spatiotemporal trajectories may be altered when structural defects are present in graphene. In this case  a small energy gap appears in the electronic band structure, which  leads to a discernible electronic effective mass $m$ \cite{advancePhys63_2014, prb93_2016}. 
In this case the free carriers are described by the relativistic Hamiltonian  \cite{epjb85_2012}
\begin{equation}
\label{eq:HwithMass}
\h = \pm v_F \sqrt{p_x^2 + p_y^2 +m^2} +V(x,y).
\end{equation}
When electrons move very fast in $x$ direction ($p_x \gg p_y$) and  the effective mass is very small ($p_x \gg m$), we expand the Hamiltonian (\ref{eq:HwithMass}) up to second order of $m/p_x$ and $p_y/p_x$ to  obtain 
\begin{equation}
\label{eq:Hm}
\h = p_x + \frac{p_y^2 +m^2}{2 p_x} +V(x,y),
\end{equation}
where  for convenience we  set $v_F=1$.
The Hamilton-Jacobi equation  of Eq. (\ref{eq:Hm}) reads 
\begin{equation}
\label{eq:HJE}
\St + \Sx +\frac{ (\Sy)^2+m^2 }{2 \Sx} +V(x,y) =0.
\end{equation}
where the classical action $S$  is defined as $p_x =\Sx $ and $p_y=\Sy$. In turn, we follow the same methodology that discussed in the main text to  calculate an approximate equation for the local curvature of the electronic flow determined through $u=\partial_{yy}S$: 
\begin{equation}
 \label{eq:ODEu_m}
\frac{d u}{dt} + \frac{u^2}{p_0+\alpha t} +\partial_{yy}V_r(t,y) = 0.
\end{equation}
We  observe that  the curvature equation (\ref{eq:ODEu_m}) is independent of  $m$ and thus,  the Dirac branched electronic flow dynamics is not affected by small defects in Dirac solids.
We infer that  Dirac branching   is robust and  not affected by the presence of few structural defects in graphene.

\section{Conclusions} 

Branching is a stochastic, spatiotemporal phenomenon that relates to the self-organization of flows in extended complex systems ranging from geophysics, to optics, to materials science and beyond.  The necessary condition for the occurrence of stochastic branching is the presence of wave-like motion in a typically weakly random environment.  The latter acts as a random index of refraction in the propagation of waves and, as a result, non-deterministic focusing and defocusing events generate wave coalescence.  The onset of branches are seen as singularities of the wave-fronts, i.e. caustics,
and as such may be
investigated analytically via effectively stochastic  differential equations. Previous work has shown that the onset and location of branching events in electronic systems scales with the statistical properties of the medium \cite{prl105_2010}.  These results are important both from the theoretical but also in practice since the onset of charge singularities is not desirable in actual devices.  

The unifying theme in the onset of branching is the existence of wave motion in a random medium.  While in geophysical or optical systems wave motion is clear, to have wave motion in materials we need to invoke wave properties of electrons or other carriers.  Electrons, as quantum particles, do have wave properties; however what is needed for branching is the collective propagation of the electrons in an organized or spatiotemporally complex flow.  This flow is induced by the collective ensemble of the electrons as they propagate in the random medium.   As has been shown clearly previously \cite{prl105_2010}, this dynamics is adequately described as a classical, non-relativistic, phase space flow.  Thus, we may conclude that branching phenomena in materials are complex phenomena induced in a corpuscular ensemble through spatial disorder and involve macroscopic focusing and singularities.

In the present work we departed from the previous analyses in material systems and focused uniquely on two dimensional novel materials such as graphene and  more generally Dirac solids.  As is well known, the electronic dispersion relation in these systems is fundamentally linear (and not quadratic, as is typically in most conventional materials) leading to effective relativistic and more specifically ultra-relativistic carrier motion. Following an approach compatible with other works in these systems
\cite{epjb85_2012} we analyzed the ultra-relativistic Hamilton-Jacobi flow problem using both analytics and exact numerics and found that also in these relativistic flows branching persists. 
In particular the ultra-relativistic caustic events occur as a result of the random particle propagation in the two dimensional random potential and that increasing the external bias, shifts the onset of the branches to latter lattice locations.  We derived a specific equation that connects the location of these events to the statistical properties of the medium as well as the bias strength.  This expression, that reduces to the previously derived law for the non-relativistic motion in the 
appropriate limit  \cite{prl105_2010} has been fully verified computationally.  A remarkable finding is that the bias field may tune the branching location and thus the phenomenon may be tested directly experimentally.   The phenomena described and predicted in this work are related to known charge puddles in graphene  \cite{prl99_2007, natPhys4_2008,  natPhys5_2009, prb91_2015, prl116_2016} and future experimental but also theoretical work in two dimensional Dirac solids should point in this direction.

%

\acknowledgements
We acknowledge support by ARO MURI Award No. W911NF-14-0247 (M.M., E.K.),  EFRI 2-DARE NSF Grant No. 1542807 (M.M.),  European Union  project NHQWAVE MSCA-RISE 691209 (G.P.T.).
We used computational resources on the Odyssey cluster of the FAS Research Computing Group at Harvard University.  M.M. and G.P.T acknowledge  helpful discussions with Dr. Ragnar Fleischmann  and Dr. Jakob J. Metzger.


\begin{thebibliography}{0}

\bibitem{natPhys12_2016}
  \Name{Degueldre H.,  Metzger J. J., Geisel T. \and Fleischmann R.}
  \REVIEW{Nature Physics}{12}{2016}{259}.
  \doi{10.1038/nphys3557}
%
\bibitem{natPhys12_2016_heller}
  \Name{Heller E.}
  \REVIEW{Nature Physics}{12}{2016}{824}.
  \doi{10.1038/nphys3558}
%
\bibitem{jfluid790_2016}
  \Name{Cousins W. \and Sapsis T. P.}
  \REVIEW{Journal of Fluid Mechanics}{790}{2016}{368-388}.
  \doi{10.1017/jfm.2016.13}
%
\bibitem{prl118_2017}
  \Name{Degueldre H., Metzger J. J., Schultheis E. \and Fleischmann R.}
  \REVIEW{Phys. Rev. Lett.}{118}{2017}{024301}.
  \doi{10.1103/PhysRevLett.118.024301}


\bibitem{wavesRandom3_1993}
  \Name{Klyatskin V.   I.}
  \REVIEW{Waves in Random Media}{3}{1993}{93-100}.
  \doi{10.1088/0959-7174/3/2/004}
%
%
\bibitem{chaos22_2012}
  \Name{Ni X.,  Lai Y.  C. \and  Wang W. X.}
  \REVIEW{Chaos}{22}{2012}{043116}.
  \doi{10.1063/1.4766757}
%
\bibitem{prl111_2013}
  \Name{Barkhofen S., Metzger J. J.  Fleischmann R. Kuhl U. \and St\"ockmann H. J.}
  \REVIEW{Phys. Rev. Lett.}{111}{2013}{183902}.
  \doi{10.1103/PhysRevLett.111.183902}
%

\bibitem{myBranchingChapter}
  \Name{Mattheakis M. \and Tsironis G. P.}
  \Book{Quodons in Mica}
    \Editor{Archilla J., Jim\'enez, N., S\'anchez-Morcillo V. \and Garc\'ia-Raffi L.}
  \Year{2015}
   \Vol{221}
  \Publ{Springer}
  \Page{425-454}
%
\bibitem{myRWs_2016}
  \Name{Mattheakis M.,  Pitsios I. J., Tsironis G. P.  \and  Tzortzakis S.}
  \REVIEW{Chaos, Solitons and Fractals}{84}{2016}{73-80}.
  \doi{10.1016/j.chaos.2016.01.008}
%

\bibitem{prl119_2017}
  \Name{Safari A., Fickler R., Padgett M. J. \and Boyd R. W.}
  \REVIEW{Phys. Rev. Lett.}{119}{2017}{203901}.
  \doi{10.1103/PhysRevLett.119.203901}


\bibitem{acoustic107_2000}
  \Name{ Wolfson M. A. \and  Tappert F. D.}
  \REVIEW{The Journal of the Acoustical Society of America}{107}{2000}{154-162}.
  \doi{10.1121/1.428297}

\bibitem{acoustic109_2001}
  \Name{ Wolfson M. A. \and  Tomsovic S.}
  \REVIEW{The Journal of the Acoustical Society of America}{109}{2001}{2693-2703}.
  \doi{10.1121/1.1362685}


\bibitem{nature410_2001}
  \Name{Topinka M. A., LeRoy B. J., Westervelt R. M., Shaw S. E. J., Fleischmann R., Heller E. J., Maranowski K. D., \and   Gossard A. C.}
  \REVIEW{Nature}{410}{2001}{183-186}.
  \doi{10.1038/35065553}

\bibitem{prl89_2002}
  \Name{Kaplan, L}
  \REVIEW{Phys. Rev. Lett.}{89}{2002}{184103}.
  \doi{10.1103/PhysRevLett.89.184103}

\bibitem{natPhys3_2007}
  \Name{Jura M. P.,   Topinka M. A.,   Urban L., Yazdani A.,  Shtrikman, H.,  Pfeiffer L. N.,  West K. W. \and  Goldhaber-Gordon D.}
  \REVIEW{Nature Physics}{3}{2007}{2007}.
  \doi{10.1038/nphys756}

\bibitem{prl105_2010}
  \Name{Metzger J. J., Fleischmann R., \and Geisel T.}
  \REVIEW{Phys. Rev. Lett.}{105}{2010}{020601}.
  \doi{10.1103/PhysRevLett.105.020601}
%
\bibitem{nature438_2005}
  \Name{Novoselov K. S., Geim A. K., Morozov S. V., Jiang D., Katsnelson M. I., Grigorieva I. V., Dubonos S. V., \and Firsov A. A.}
  \REVIEW{Nature}{438}{2005}{197-200}.
  \doi{10.1038/nature04233}

\bibitem{advancePhys63_2014}
  \Name{Wehling T.O.,  Black-Schaffer A.M., \and  Balatsky A.V.}
  \REVIEW{Advances in Physics}{63}{2014}{1-76}.
  \doi{10.1080/00018732.2014.927109}

\bibitem{prb93_2016}
  \Name{Kumar A., Nemilentsau A., Fung K. H., Hanson G. F., Nicholas X. \and Low T.}
  \REVIEW{Phys. Rev. B}{93}{2016}{041413}.
  \doi{10.1103/PhysRevB.93.041413}

\bibitem{nanoLet16_2016}
  \Name{Bhandari S., Lee G. H., Klales A., Watanabe K., Taniguchi T., Heller E.,  Kim P., \and Westervelt R. M.}
  \REVIEW{Nano Letters}{16}{2016}{1690-1694}.
  \doi{10.1021/acs.nanolett.5b04609}

\bibitem{science351_2016}
  \Name{Crossno J.  \etal}
  \REVIEW{Science}{351}{2016}{1059-1061}.
  \doi{10.1126/science.aad0343}

%
\bibitem{epjb85_2012}
  \Name{Pototsky A., Marchesoni F., Kusmartsev F. V.  H{\"a}nggi P. \and Savelev S. E.}
  \REVIEW{The European Physical Journal B}{85}{2012}{356}.
  \doi{10.1140/epjb/e2012-30716-7}

\bibitem{pre87_2013}
  \Name{Pototsky A. \and Marchesoni F.}
  \REVIEW{Phys. Rev. E}{87}{2013}{032132}.
  \doi{10.1103/PhysRevE.87.032132}
%

\bibitem{natPhys4_2008}
  \Name{Martin J., Akerman N., Ulbricht G., Lohmann T., Smet J. H., von Klitzing K.,  \and Yacoby, A.}
  \REVIEW{Nature Physics}{4}{2008}{144}.
  \doi{10.1038/nphys781}

\bibitem{prl99_2007}
  \Name{Tan Y. W., Zhang Y., Bolotin K., Zhao Y., Adam S., Hwang E. H., Das Sarma S., Stormer H. L. \and Kim P.}
  \REVIEW{Phys. Rev. Lett.}{99}{2007}{246803}.
  \doi{10.1103/PhysRevLett.99.246803}

\bibitem{natPhys5_2009}
  \Name{Zhang Y., Brar V. W., Girit C.,  Zettl A. \and Crommie M. F.}
  \REVIEW{Nature Physics}{5}{2009}{722}.
  \doi{10.1038/nphys1365}

\bibitem{prb91_2015}
  \Name{Martin S. C.,  Samaddar S.,  Sac\'ep\'e B.,  Kimouche A.,  Coraux J.,  Fuchs F.,  Gr\'evin B.,  Courtois H. \and Winkelmann C. B.}
  \REVIEW{Phys. Rev. B}{91}{2015}{041406}.
  \doi{10.1103/PhysRevB.91.041406}

\bibitem{prl116_2016}
  \Name{Samaddar S.,  Yudhistira I.,  Adam S.,  Courtois H. \and Winkelmann C. B.}
  \REVIEW{Phys. Rev. Lett.}{116}{2016}{126804}.
  \doi{10.1103/PhysRevLett.116.126804}

%

\end{thebibliography}

\end{document}